\documentclass[a4paper]{article}
\usepackage[a4paper, left=30mm, right=30mm, top=30mm, bottom=30mm]{geometry}
\usepackage[T1]{fontenc}
\usepackage[utf8]{inputenc}
\usepackage{iwona}
\usepackage[english]{babel}
\usepackage{graphicx}
\usepackage[hidelinks]{hyperref}
\begin{document}

\hspace{2in}
\begin{center}
\begin{minipage}{5in}
\begin{center}
	{\Large Visualizing signatures of human activity in cities across the globe} \\[1ex]
		D\'aniel Kondor$^{1,2,\ast}$, Pierrick Thebault$^{1}$, Sebastian Grauwin$^{1}$, Istv\'an G\'odor$^{2}$, \\%
		Simon Moritz$^{3}$, Stanislav Sobolevsky$^{1}$, Carlo Ratti$^{1}$
\end{center}
	{\bf 1} SENSEable City Laboratory, Massachusetts Institute of Technology, Cambridge, MA, USA \\
	{\bf 2} Ericsson Research, Budapest, Hungary \\
	{\bf 3} Ericsson Research, Sweden \\
	$\ast$ E-mail: dkondor@mit.edu \\[5ex]
	
\begin{center}
\includegraphics[width=5in]{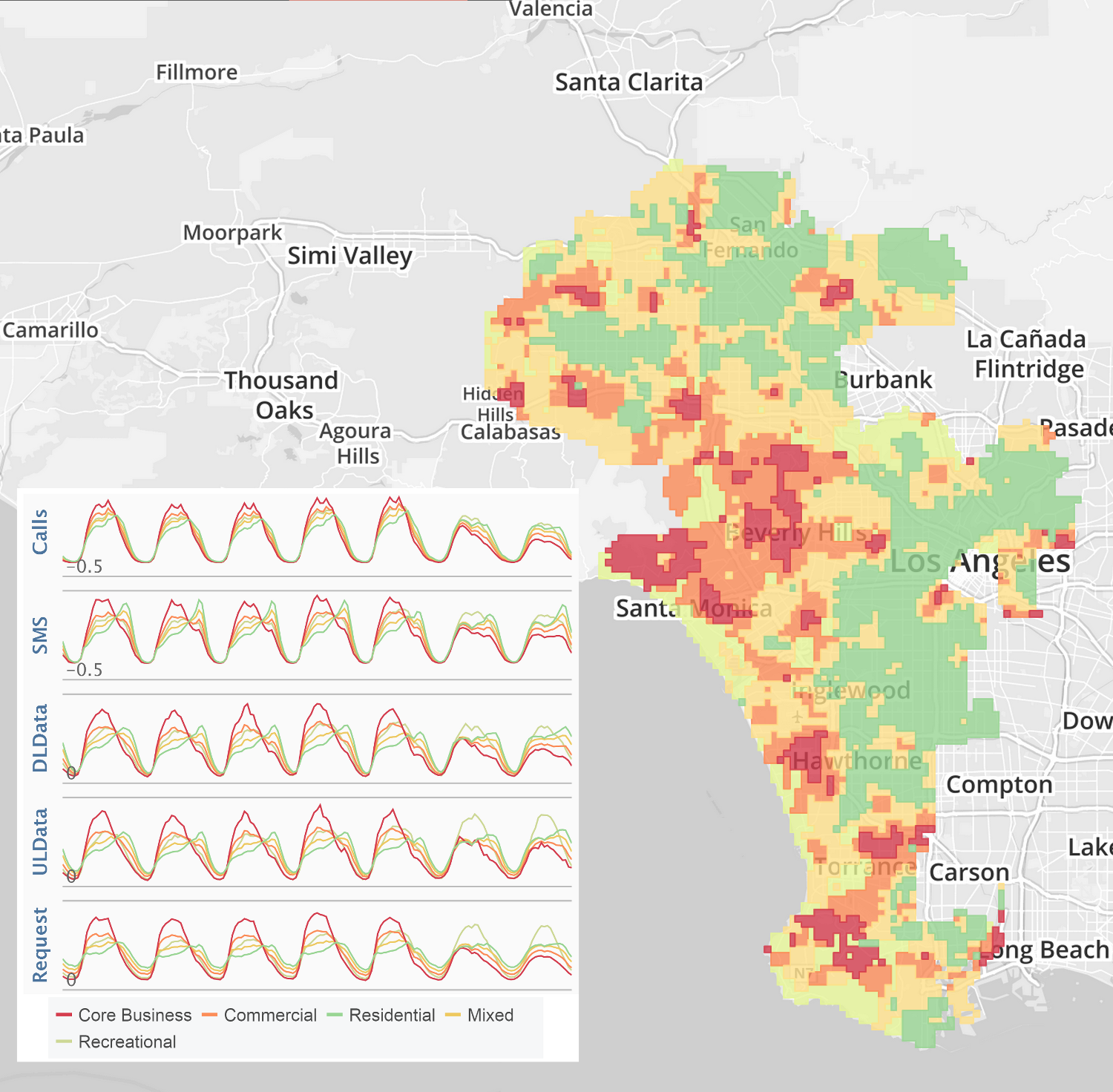}\\
\small Functional clusters found in Los Angeles based on typical telecommunications activity, visualized on the interactive platform
\url{http://manycities.org}.
\end{center}
\end{minipage}
\end{center}

\newpage

\begin{abstract}
	The availability of big data on human activity is currently changing the way we look at our surroundings. With the high penetration of mobile phones, nearly everyone is already carrying a high-precision sensor providing an opportunity to monitor and analyze the dynamics of human movement on unprecedented scales. In this article, we present a technique and visualization tool which uses aggregated activity measures of mobile networks to gain information about human activity shaping the structure of the cities. Based on ten months of mobile network data, activity patterns can be compared through time and space to unravel the ``city's pulse'' as seen through the specific signatures of different locations. Furthermore, the tool allows classifying the neighborhoods into functional clusters based on the timeline of human activity, providing valuable insights on the actual land use patterns within the city. This way, the approach and the tool provide new ways of looking at the city structure from historical perspective and potentially also in real-time based on dynamic up-to-date records of human behavior. The online tool presents results for four global cities: New York, London, Hong Kong and Los Angeles.
\end{abstract}

Increasing dynamics of urban transformation in a rapidly developing world calls for adaptive urban planning. One of the central questions for city planners is to know the function of different neighborhoods, which is crucial for understanding their needs. Traditionally, cities maintain a database of land-use classification, which is based on official records. While such data is generally of very high resolution, its nature limits the ability to follow changes dynamically. But today's challenges come together with new solutions to them: the use of anonymized mobile phone data enables dynamic measurement of human activity and sensing the unique signatures of each neighborhood of a city based on that. Exploiting these new possibilities, recent years saw increasing usage of mobile phone data for better understanding people's presence around the city~\cite{ref1,ref2}, human mobility~\cite{ref3,ref4,ref5}, structure of the city~\cite{ref6,ref7,ref8} and many other applications, including regional delineation~\cite{ref9,ref10}.

% \the\textwidth
\begin{figure}[h]
	\centering
	\includegraphics{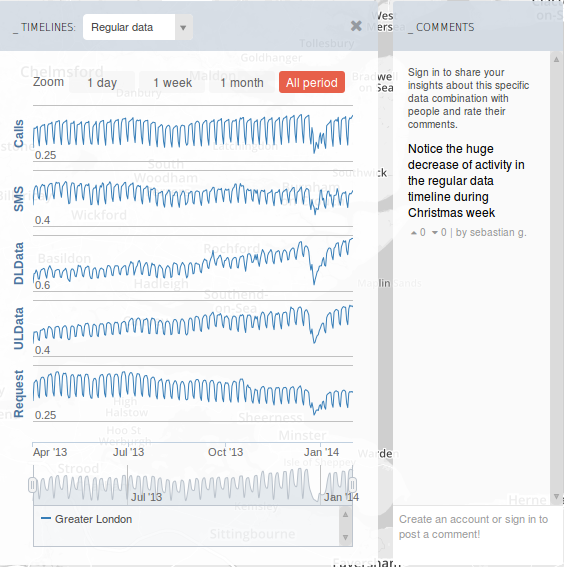}
	\includegraphics{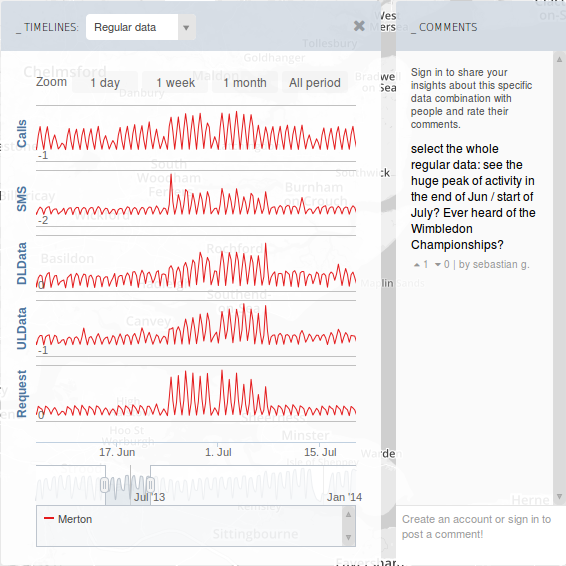}
	\includegraphics{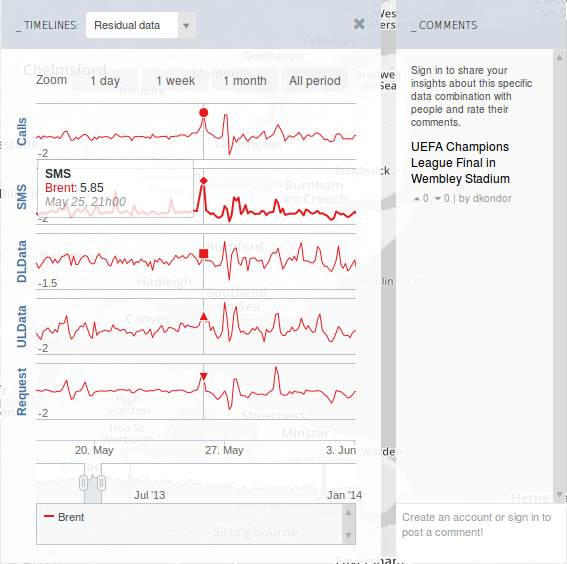}
	\caption{Identifying long-term trends and events in mobile network activity in the Greater London area. Left: city-wide aggregated data shows a steady increase in data traffic; also notice the large drop around Christmas. Middle: Concentrating on the Merton district, the effect of the Wimbledon Tennis Championship can be seen in network activity. Right: Looking at the residual data, special events can be identified by large peaks. The example given here is the district containing WembleyStadium which is a major sports venue in London; the timeline is focused on the date of the 2013 UEFA Champions League Final.}
\end{figure}

The presented tool~\cite{ref11}, developed in collaboration between MIT SENSEable City Laboratory and Ericsson, allows the comparative analysis of city structure with the use of aggregated mobile network activity data in four global cities: New York, London, Hong Kong and Los Angeles. Further research insights coming from such analysis are presented in~\cite{ref12}. The tool uses aggregated data collected between April 2013 and January 2014 in the four cities provided by mobile network operators with representative market shares. The data includes measurements of human phone activity aggregated on the level of mobile antennas within 15-minute time intervals. These include the number of calls placed and text messages sent, the amount of data downloaded and uploaded, and the number of data requests (number of individual times data transfer was initiated) for each antenna in each 15-minute time window. This way, the aggregated data does not include any sensitive customer information, but provides enough detail about the typical usage patterns on the scale of small neighborhoods. We proceed by further spatial aggregation so as to account for the variation of antenna-level activity volumes due to technical constraints of cellular network operation. For most of the analysis, we normalize the resulting time series of different locations; we expect that the most information can be gained from comparing the shape of time series, i.e.~how activity is distributed over a day or week. Looking at the spatial distribution of activity volumes is possible in a separate usage mode in the tool.

\begin{figure}
	\centering
	\includegraphics{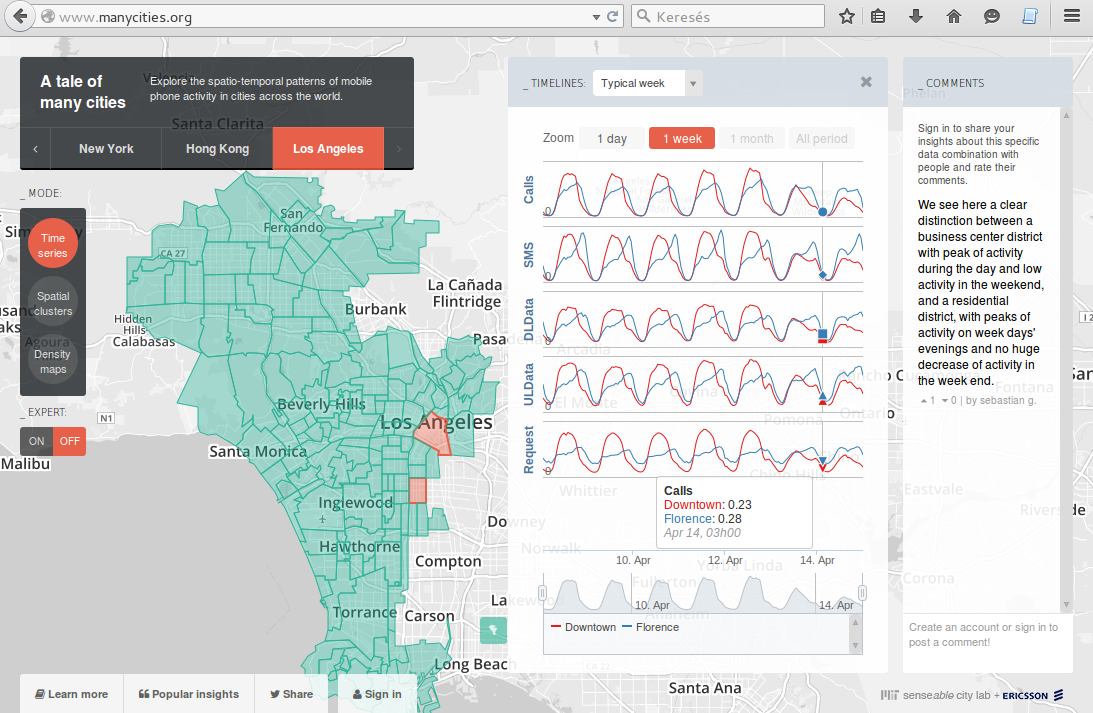}
	\caption{Comparing the activity profile of the typical week, there is a significant distinction between a central business area (downtown Los Angeles, the red curve) and a residential neighborhood (Florence, in blue).}
\end{figure}

Possible usage of the data is demonstrated via three possible usage modes in the online tool~\cite{ref11}: time series, spatial clusters and density maps. The first, time series mode facilitates the exploration of human activity measures on the level of city districts; possible usage is illustrated in Figs.~1, 2 and~3. In this mode, one can look at individual time series throughout the measurement period making possible to identify long-term trends and special events, and weekly averages enabling the comparison of typical activity patterns in different neighborhoods or in different cities. Notable features in long-term time series include the steady increase in data traffic, the distinct effect of holidays, and specific signatures of some important events like the Wimbledon tennis championship or football matches in London. Looking at typical week time series, patterns representative of cities' business centers and residential areas can be easily identified. Apart from looking at the long-term time series of activity and weekly averages, it is also possible to inspect the residual activity, i.e.~difference of activity on a specific week from the average. Special events can also be noticed by peaks in these time series as illustrated by the example of Wembley Stadium in Fig.~1.

\begin{figure}
	\centering
	\includegraphics{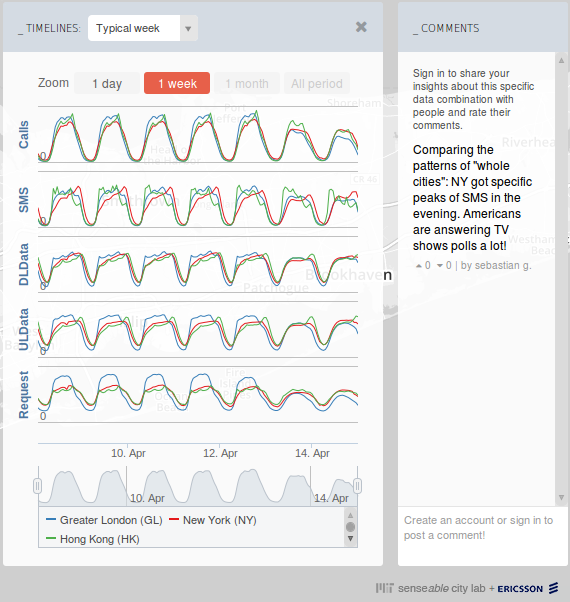}
	\caption{Comparing city-wide aggregated activity profiles in three cities. Apart from showing a high degree of similarity corresponding to the natural circadian rhythm of humans, there are notable differences, e.g.~text message activity peaks in the morning in Hong Kong, in the evening in New York and midday in London or the abrupt decrease of data traffic in London when contrasting the evening peaks in the other two cities. We speculate that the latter observationis the effect of cellular data traffic being especially expensive in London, prompting people to switch to much cheaper wifi networks when at home.}
\end{figure}

The second, spatial clusters mode of the online tool allows the interactive exploration of results on decomposing the cities into functional clusters based on the mobile network activity time series. This analysis was performed on the typical week time series of network activity aggregated into a regular grid. Each of the functional clusters can then be represented with a cluster-wide average time series and a map displaying the grid pixels which belong to it. As an example, the functional clusters of New York City are displayed in Fig.~4. Looking at the average cluster time series, we can classify them as e.g. core business, where activity is especially high in normal working hours and low otherwise; residential, where the peak of daily activities is in the evenings; leisure and parks, where activity has noticeable peaks in weekends during the day. It is also possible to compare clusters from different cities. For example, in Fig.~4, we show comparison of clusters identified as core business and residential in New York and Hong Kong. It is remarkable that the activity signatures of the two business clusters seem much more similar than the signatures of the two residential clusters. A more detailed analysis of results of clustering, including in-depth discussion of the interpretation of the clusters, comparison of results with traditional, census-based data, and a systematic analysis of differences among the four cities is presented in our related scientific publication~\cite{ref12}.

\begin{figure}
	\centering
	\includegraphics{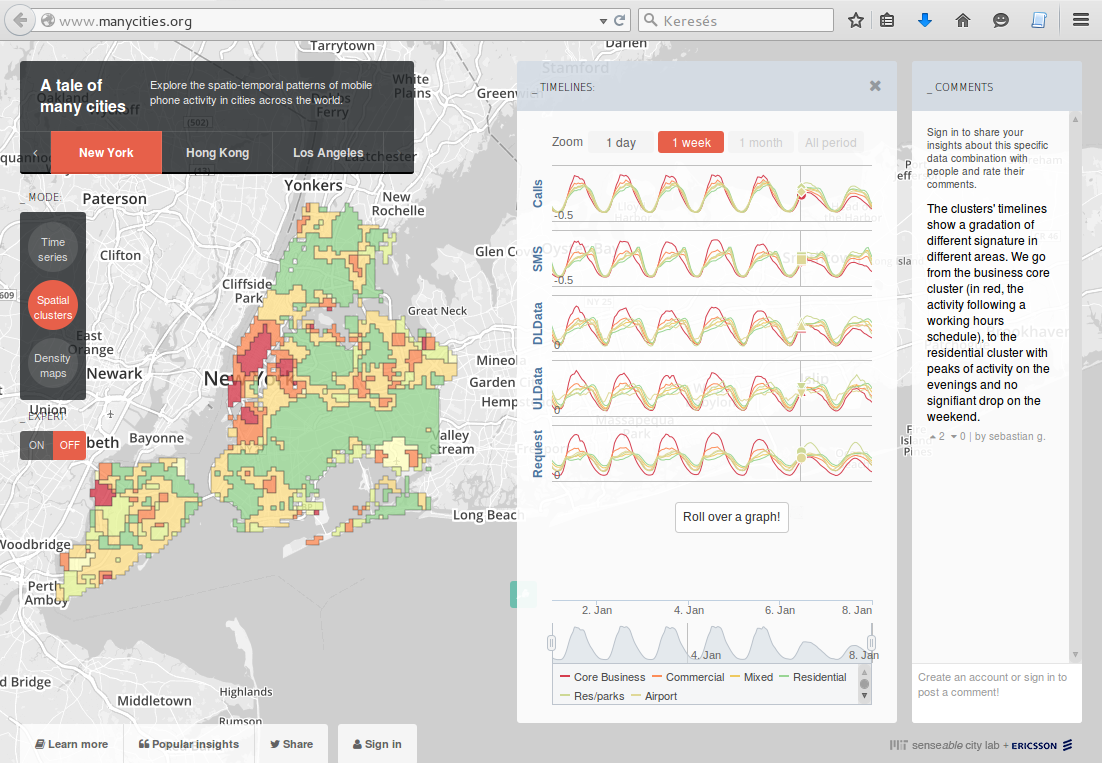}
	\includegraphics{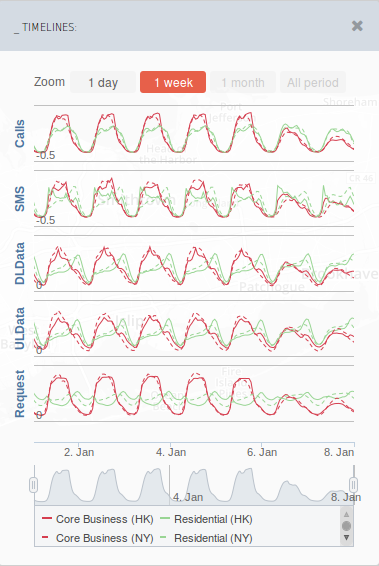}
	\caption{Functional clusters formed on the similarity of typical activity time series uncover important structural characteristics of cities. Left: Functional clusters found in New York; Middle: typical time series of the clusters; Right: Comparing business and residential clusters in New York with the same in Hong Kong.}
\end{figure}

The third mode of the online tool presents the spatial distribution of activity volumes in the cities. It also allows inspecting the ratio of volumes of different activity types, giving an overview of which type of activity is preferred in different parts of the cities. An example is displayed in Fig.~5, showing the spatial distribution of call volumes in Hong Kong.

\begin{figure}
	\centering
	\includegraphics{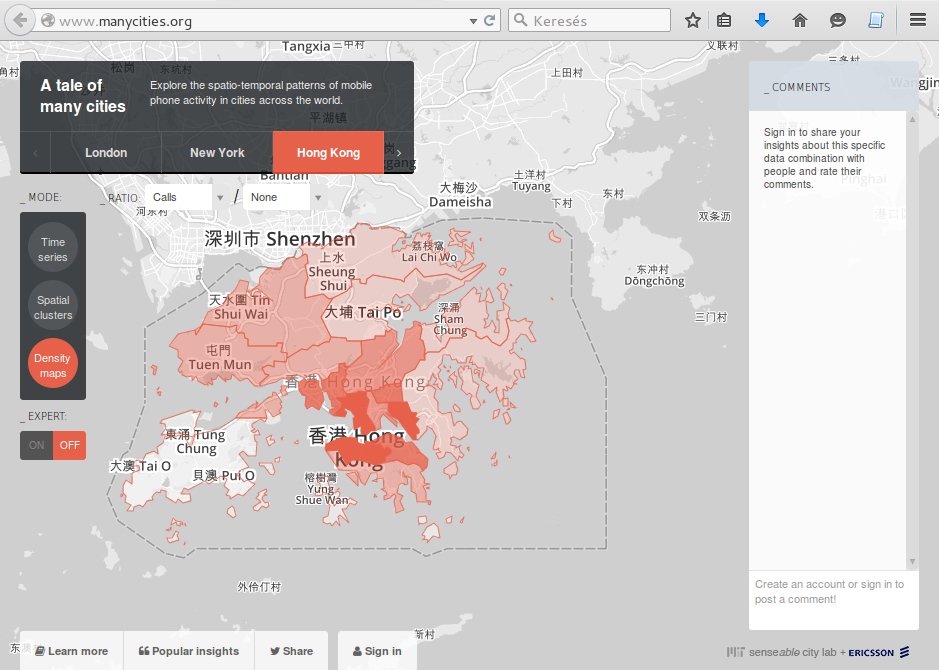}
	\caption{Spatial distribution of average volume of calls placed in Hong Kong.}
\end{figure}

We believe the presented tool to be a valuable supplement for urban planning decision support as well as for providing important insights on the way people use the city for researchers, urban stakeholders and general public.

\section*{Acknowledgements}
The authors wish to thank Ericsson for providing the aggregated phone activity records. We also thank Zsófia Kallus at Ericsson Research for stimulating discussions. We further thank Ericsson, MIT SMART Program, Accenture, Air Liquide, BBVA, The Coca Cola Company, Emirates Integrated Telecommunications Company, The ENEL foundation, Expo 2015, Ferrovial, Liberty Mutual, The Regional Municipality of Wood Buffalo, Volkswagen Electronics Research Lab, UBER and all the members of the MIT Senseable City Lab Consortium for supporting the research.

\end{document}